\documentclass{aa}  
\usepackage{graphicx}
\usepackage{natbib}
\usepackage{txfonts}

\def\ms{\hbox{\,m\,s$^{-1}$}}         
\def\m2s2{\hbox{\,m$^{2}$\,s$^{-2}$}} 
\def\kms{\hbox{\,km\,s$^{-1}$}}       
\def\vsini{\hbox{$v$\,sin\,$i$}}      
\def\sini{\hbox{sin\,$i$}}      

\def\Msun{\hbox{$M_{\odot}$}}             
\def\Mjup{\hbox{$\mathrm{M}_{\rm Jup}$}}

\def \1s{$1\,\sigma$}
\def \kid{$\chi^2$}
\def \t0{T$_0$}

\def \sophie{{\it SOPHIE}}

\def\logrhk{$\log R'_\mathrm{HK}$}

\begin{document} 

\title{The $SOPHIE$ search for northern extrasolar planets\thanks{Based on observations made with 
$SOPHIE$ spectrograph on the 1.93-m telescope at Observatoire de Haute-Provence (CNRS / OSU Pyth\'eas), 
France (program 07A.PNP.CONS). }}

   \subtitle{VIII. A warm Neptune orbiting HD164595}

\author{
Courcol, B. \inst{1}
Bouchy, F. \inst{1}
\and Pepe, F. \inst{2}
\and Santerne, A. \inst{3}
\and Delfosse, X. \inst{4,10}
\and Arnold, L. \inst{6}
\and Astudillo-Defru, N. \inst{4,10}
\and Boisse, I. \inst{1}
\and Bonfils, X. \inst{4,10}
\and Borgniet, S. \inst{4}
\and Bourrier, V. \inst{2}
\and Cabrera, N. \inst{4,10}
\and Deleuil, M. \inst{1}
\and Demangeon, O. \inst{1}
\and D\'iaz, R.F. \inst{1,2}
\and Ehrenreich, D. \inst{2}
\and Forveille, T. \inst{4,10} 
\and H\'ebrard, G. \inst{5,6}
\and Lagrange, A.M. \inst{4}
\and Montagnier, G. \inst{5,6}
\and Moutou, C. \inst{1,8} 
\and Rey, J. \inst{2}
\and Santos, N.C. \inst{3,7}
\and S\'egransan, D. \inst{2}
\and Udry, S. \inst{2}
\and Wilson, Paul A. \inst{5,9}
}

\institute{
Aix Marseille University, CNRS, Laboratoire d'Astrophysique de Marseille UMR 7326, 13388 Marseille cedex 13, France\\
bastien.courcol@lam.fr
\and
Département d'Astronomie de l'Université de Genève, 51. ch. des Maillettes - Sauverny, CH-1290 Versoix, Switzerland
\and  
Instituto de Astrofísica e Ciências do Espaço, Universidade do Porto, CAUP, Rua das Estrelas, 4150-762 Porto, Portugal
\and
Univ. Grenoble Alpes, IPAG, F-38000 Grenoble, France
\and
Institut d'Astrophysique de Paris, UMR7095 CNRS, Universit\'e Pierre \& Marie Curie, 98bis boulevard Arago, 75014 Paris, France
\and
Aix Marseille Universit\'e, CNRS, OHP, Institut Pyth\'eas UMS 3470, 04870 Saint-Michel-l'Observatoire, France
\and
Departamento de Física e Astronomia, Faculdade de Ciências, Universidade do Porto, 4169-007 Porto, Portugal
\and
CFHT, 65-1238 Mamalahoa Hwy, 96743 Kamuela HI, USA
\and 
Sorbonne Universités, UPMC Univ Paris 06, UMR 7095, Institut d’Astrophysique de Paris, F-75014, Paris, France
\and
CNRS, IPAG, F-38000 Grenoble, France
}

   \date{Received 16/04/2015; accepted 12/06/2015}
 
 \abstract
 {High-precision radial velocity surveys explore the population of low-mass exoplanets orbiting bright stars. This allows accurately
deriving their orbital parameters such as their occurrence rate and the statistical distribution of their properties. Based on
this, models of planetary formation and evolution can be constrained. The {\sophie} spectrograph has been continuously improved in
past years, and thanks to an appropriate correction of systematic instrumental drift, it is now reaching 2 {\ms} precision in radial velocity measurements on all timescales. As part of a dedicated radial velocity survey devoted to search for low-mass planets around a sample of 190 bright solar-type stars in the northern hemisphere, we report the detection of a warm Neptune with a minimum mass of $16.1 \pm 2.7$ M$_{\oplus}$ orbiting the solar analog HD164595 in $40 \pm
0.24$ days . We also revised the parameters of the multiplanetary system around HD190360. We discuss this new detection in the context of the upcoming space mission CHEOPS, which is  devoted to a transit search of bright stars harboring known exoplanets.}


   \maketitle
%

\section{Introduction}

    Over the past 20 years, the radial velocity (RV) technique has benefited from continuous developments. Several major milestones in detection and characterization of extrasolar planets have been made possible with this technique, either in the context of large surveys or in follow-ups of transiting candidates. These developments allowed a few stabilized spectrographs like HARPS \citep{Mayor2003} to gradually scan the area of Neptune-like objects and super-Earths and detect exoplanets with minimum masses as low as 1 M$_\oplus$ \citep{Dumusque2012}. 
    
Recent exoplanet RV surveys show that over 40\% of non-active solar-type stars probably harbor low-mass exoplanets \citep{Mayor2011,Howard2010} with a mass distribution that seems to increase sharply toward low masses. In addition, most of the low-mass exoplanets belong to multiple compact systems (e.g., \citealt{Lovis2011}). These results overall agree with the observed properties of the transiting exoplanets detected by the Kepler Mission \citep{Borucki2011,Howard2012,Marcy2014}. Detection and characterization of new low-mass exoplanets orbiting bright stars is motivated by the need to obtain accurate measurements of the projected mass and the eccentricity and to identify all the components of multiple systems to constrain models of exoplanet
evolution. Such exoplanets orbiting bright stars will also be key targets for a deeper and more extended characterization with CHEOPS \citep{Broeg2013,Fortier2014}, for instance, which searches for the transit, and JWST \citep{Beichman2014}, which searches for atmospheric signatures.        

The {\sophie} spectrograph \citep{Perruchot2008} on the 1.93
m telescope of the Observatoire de Haute Provence was improved in June 2011 with the implementation of octagonal-section fibers \citep{Perruchot2011}. This optimization led to an improvement in RV precision to nearly 2 {\ms} \citep{Bouchy2013} on timescales of a few tens of days. This precision level is critical for detecting low-mass exoplanets around solar-type stars. In the context of the subprogram ``{\it High precision search for Neptunes and super-Earths}'' of the {\sophie} extrasolar planets research consortium \citep{Bouchy2009}, we redefined a sample of 190 bright stars on the following criteria : 1) in the northern hemisphere, 2) with spectral type G and K and $0.6 \le B-V \le 1.4 $, 3) on a volume limited to 35 pc, 4) non-active (vsini $\le$4.5 {\kms}, $\log{R'_\mathrm{HK}}$ $\le$ -4.8), 5) not part of a binary system and not harboring giant planets at short orbital period, and 6) not part of RV surveys carried out with HARPS-N GTO \citep{Cosentino2012} at the TNG. This subprogram is intensively conducted since June of 2011.

We here report on the detection of a Neptune-like exoplanet candidate orbiting the solar analog HD164595. Section 2 describes the SOPHIE observations and the data reduction. In Sect. 3 we describe the correction of the systematic instrumental drift necessary to reach the 2 {\ms} precision level. This correction is then validated in Sect. 4 using the known planetary system HD190360, for which we provide an updated orbital solution. Section 5 details our data analysis of HD164595 and summarizes the orbital and physical properties of the planet. We finally present our conclusions in Sect. 6.

\section{Observations and data reduction}

Spectroscopic observations were conducted with the {\sophie} spectrograph \citep{Perruchot2008,Perruchot2011}.  {\sophie} is a fiber-fed environmentally stabilized echelle spectrograph covering the visible range from 387 to 694 nm. We used the high-resolution (HR) mode ($\lambda / \Delta\lambda$ =75,000) with thorium-argon simultaneous wavelength calibration on fiber B.

In total we gathered 75 measurements of HD164595 with {\sophie} over 2.2 years. The exposure time was set to between 800 and 1200 seconds to reach a signal-to-noise ratio (S/N) of 140 at typical photon-noise RV uncertainties below 1.0 {\ms} while simultaneously averaging p-mode stellar oscillations. Taking into account the wavelength calibration uncertainty (close to 1.0 {\ms}), the average RV uncertainty is 1.43 {\ms}. Nine spectra were identified as low quality. One measurement has a S/N that is half of the average, indicating a strong atmospheric absorption. Three measurements are affected by moon light contamination, which cannot be corrected with the simultaneous thorium mode. Five measurements present a flux anomaly of the thorium lamp, which prevented us from computing the simultaneous drift. They were discarded for the data analysis in Sect. 4.4.

The spectra were reduced and extracted using the {\sophie} pipeline \citep{Bouchy2009}, and the resulting wavelength-calibrated 2D spectra were correlated using a numerical binary mask corresponding to spectral type G2 to obtain the radial velocity measurement \citep{Baranne1996,Pepe2002}. The full width at half maximum (FWHM), contrast,  and bisector span of the cross-correlation function (CCF) are also provided by the pipeline.

The complete radial velocity dataset, after the zero-point drift correction described in the next section, is reported in the
Appendix and plotted in Fig. \ref{HD164595_orb}. 


\section{Correction of the systematic instrumental drift}

\begin{figure}
  \centering
  \includegraphics[width=\hsize]{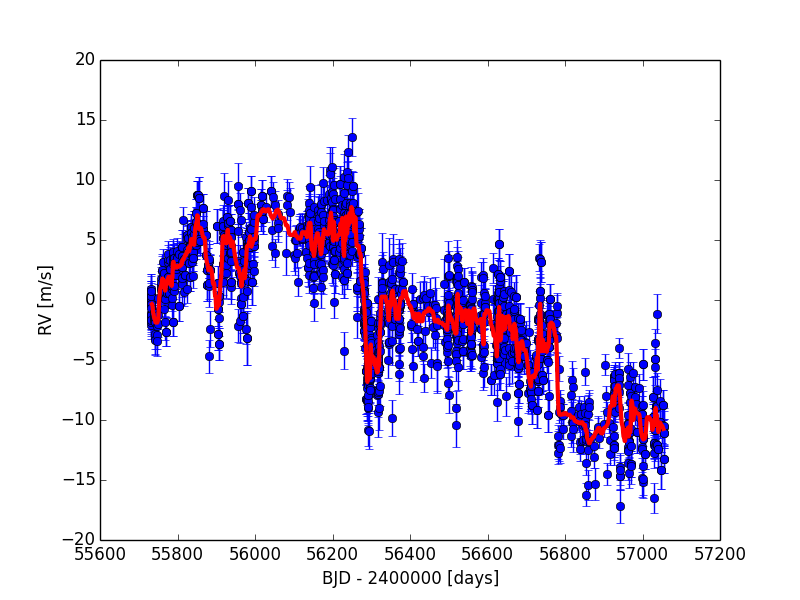}
   \includegraphics[width=\hsize]{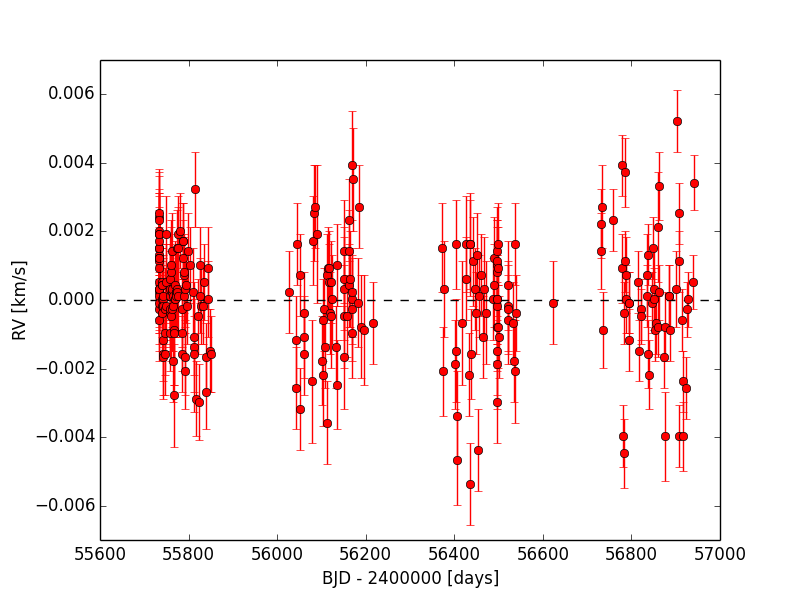}
      \caption{\textit{Upper panel} : Zero-point drift correction (red line) as a function of time and the adjusted RV data set used to create it (blue dots). \textit{Bottom panel} : RV time series of HD185144 after the zero-point drift correction. This correction decreases the RMS from 5.37 {\ms}  to 1.61 {\ms} .}
         \label{constantes}
\end{figure}

Despite a clear improvement in RV precision provided by the implementation of octogonal fibers on timescales of a few tens of days \citep{Bouchy2013,Perruchot2011}, we identified long-term variations of the zero point of the instrument that added up to about $\pm$ 10 {\ms} over the last 3.5 years (cf. Fig. \ref{constantes}, upper panel). The first ramp seems to be correlated with the thorium-argon lamp aging. One jump of about -7 m/s is related to the implementation of additional octagonal fibers after the double scrambling in December 2012 (BJD=55950). A second jump of -7 m/s at BJD=56775 appears after the coating of the secondary mirror of the 1.93 m telescope, which introduced a significant change in the flux balance across the spectral range. Other events, with a timescale of a few weeks, are clearly correlated with strong changes of the outside temperature, which are propagated to the tripod of the spectrograph at a level of a few tenths of degrees through the telescope pillar. A thermal control loop of the spectrograph tripod is under development to avoid such thermal conduction coming from the telescope pillar.       

To track this long-term zero-point drift of the instrument, we systematically monitored a set of RV constant stars
each night, namely HD185144 \citep{Howard2010}, HD9407, HD221354, and HD89269A \citep{Howard2011}, which were monitored with HIRES on the 10.2m Keck telescope with an RV dispersion of 2.0, 1.7, 1.9, and 2.0 {\ms} , respectively, over several years. In addition to this set of four RV constant stars, we also used 51 targets from our sample that have at least ten measurements. We recursively built an RV constant master using these two sets of stars. A first master was created using a running median on the four reliable constant stars. This master was then subtracted from the second group, and all the stars with a corrected RMS lower than a threshold of 3 {\ms} were included in the first subset. The offset for each star was adjusted as well to minimize the dispersion of the residuals. A double weight was given to the four reliable constant stars. This process was repeated until convergence, that is, when no additional star was added to the first subset. 

The running median was implemented with the criterium to average 15 measurements. This was preferred instead of taking a fixed time window to give the same weight to all correction points. Consequently, a period with a higher density of observations will increase the temporal resolution of the correction. On the other hand, a period with sparse measurements will lead to a lower temporal resolution instead of a poorer constraint on the correction point. In practice, with this criteria of 15 measurements averaged, the typical timescale of the correction is nine days.  

The most satisfying correction uses a threshold of 3 {\ms} and converges after four iterations with a total of 23 constant stars in the final set. The radial velocities of these 23 constant stars as well as the derived RV constant master \footnote{The RV constant master is available upon request.} are displayed in the top panel of Fig. \ref{constantes}. RV variations are clearly visible on different timescales, from a few days to a global trend over the last 3.5 years. This approach allowed us to gather enough data to correctly cover the time series and to average all physical effects such as stellar activity or signals of small planets, leaving only the instrumental systematics.

\begin{figure}
   \centering
   \includegraphics[width=\hsize]{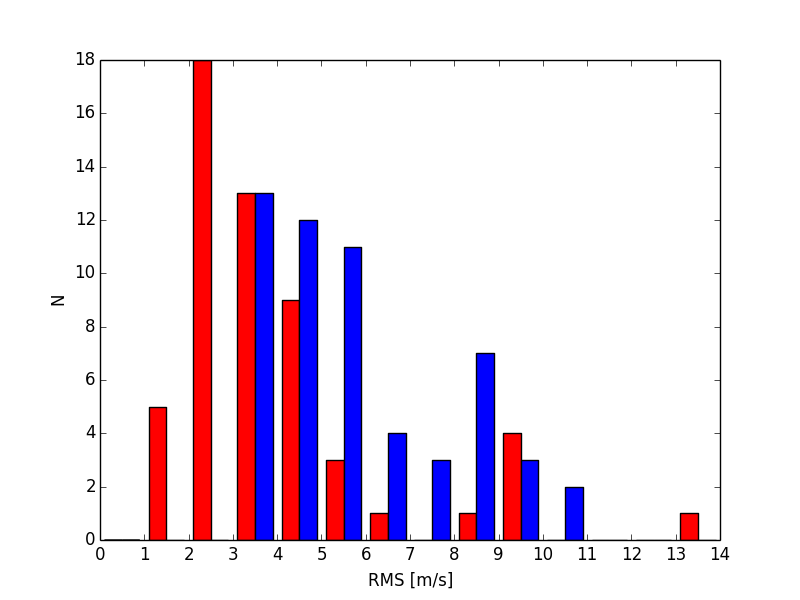}
      \caption{Initial (blue) and final (red) distribution of the RMS in the 55-star sample.}
         \label{histo}
\end{figure}

To illustrate the gain provided by our correction of the systematic instrumental drift, the RV of the constant star HD185144 is displayed in the bottom panel of Fig. \ref{constantes}. For this target, the RV dispersion decreases from 5.4 to 1.6 {\ms} when the RV constant master is subtracted. Figure \ref{histo} shows the RMS distribution of the 55 stars (including the four reliable constant stars) before and after correction of the RV constant master 
and illustrates the fact that the {\sophie} RV precision is close to 2~ {\ms}. \\

\section{Effect on the planetary system HD190360}

\subsection{Validation of the correction}

To validate the correction of the RV constant master, we analyzed the known planetary system HD190360 that was observed with {\sophie} in the context of the science validation of the octogonal fibers \citep{Bouchy2013}. The planetary system HD190360 is composed of a long-period Jupiter \citep{Naef2003} and a warm Neptune \citep{Vogt2005}. A total of 41 measurements of HD190360 were gathered with SOPHIE using the same setting as for HD164595. The radial velocity table is reported in the Appendix. The RV dispersion is 9.5 {\ms} and 9.3 {\ms} before and after correction of the RV constant master, respectively.  The span of our observations
was 2.4 years, which is about one-eighth of the period of HD190360b.
This does not allow efficiently constraining the orbital parameters of this planet. We therefore focus our analysis on the warm Neptune HD190360c.

HD190360 is a quiet star with a {\logrhk} derived from SOPHIE spectra of -5.13.  No periodic signals in either the bisector, the CCF-FWHM, or the activity index are present at the two periods found in RVs, which excludes stellar activity as the origin of the Keplerian signals. 

To study the impact of our zero point drift correction, we analyzed both the HD190360 corrected and uncorrected data set with the same methodology. We used the planet analysis and small transit investigation software (PASTIS) \citep{Diaz2014}, which is a Bayesian analysis software, for fitting a two-planet model. HD190360b orbit was not fixed but subjected to strong priors for the Bayesian analysis. Normal distributions were used, centered on the published parameters with widths equal to the published errors in \cite{Wright2009}. On the other hand, the other planet parameters were let free, with a Jeffrey distribution for the period, a beta distribution for the eccentricity as defined in \cite{Kipping2013}, and uniform distributions for the other parameters. We ran 20 MCMC chains with 300 000 steps each. Only the stationary parts of each chain were kept. The distributions of parameter values of all the remaining uncorrelated chain links then correspond to the target joint posterior-probability distributions for the model parameters.

Table \ref{table:HD190360_SOPHIE} reports our orbital parameters and their uncertainties for both cases. The results are very similar for the two data sets and agree well (within $1\sigma$) with \cite{Wright2009}. However, the errors in the uncorrected data are systematically higher (by up to 50\%). The dispersion of the residuals is better in the corrected set with 2.1 {\ms}, which is consistent with the precision obtained on the constant stars, instead of 3.5{\ms} with the uncorrected data. Moreover, the additional jitter necessary to set the reduced {\kid} to 1 is 1.9 {\ms}, instead of 2.9 {\ms} for the uncorrected data set. This test demonstrates the relevance of the zero-point drift correction, which significantly improves the quality of the RV data. 

\begin{table}
\caption{Comparison of the orbital solutions for HD190360c with corrected and uncorrected SOPHIE observations.}             
\label{table:HD190360_SOPHIE}      
\centering                          
\begin{tabular}{c c c}        
\hline\hline                 
Parameter & HD190360c & HD190360c \\    
 & (uncorrected data) & (corrected data) \\
\hline                        
   P [days] & $17.117 \pm 0.015$ & $17.127 \pm 0.011$\\      
   $T_0$ [BJD] & $56149 \pm 4$  & $56134 \pm 4$ \\
   $e$ & $0.088 \pm 0.1$ & $0.062^{+0.086}_{-0.047}$  \\
   $\omega$ [deg] & $237.1 \pm 94$  & $248.5^{+80}_{-210}$\\
   K [\ms] & $5.72 \pm 0.79$ & $5.64 \pm 0.55$\\
\hline
   $N_{meas}$ & 41 & 41 \\      
   $\sigma$ (O - C) [\ms] & 2.9 & 2.1 \\   
   Jitter (O - C) [\ms] & 3.5 & 1.9 \\   
\hline                                   
\end{tabular}
\end{table}

\subsection{Analysis of the joint HIRES and SOPHIE data}

\begin{figure}
   \centering
   \includegraphics[width=\hsize]{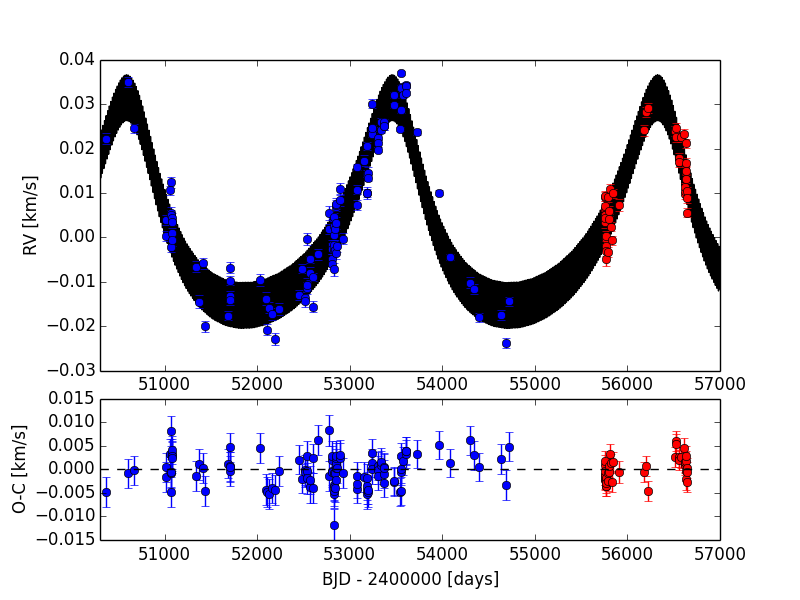}
      \caption{Radial velocities of HD190360 for HIRES (blue dots) and SOPHIE (red dots) with the best-fit orbital solution (solid line). Error bars include the photon noise and jitter.}
         \label{HD190360tot_orb}
\end{figure}

\begin{figure}
   \centering
   \includegraphics[width=\hsize]{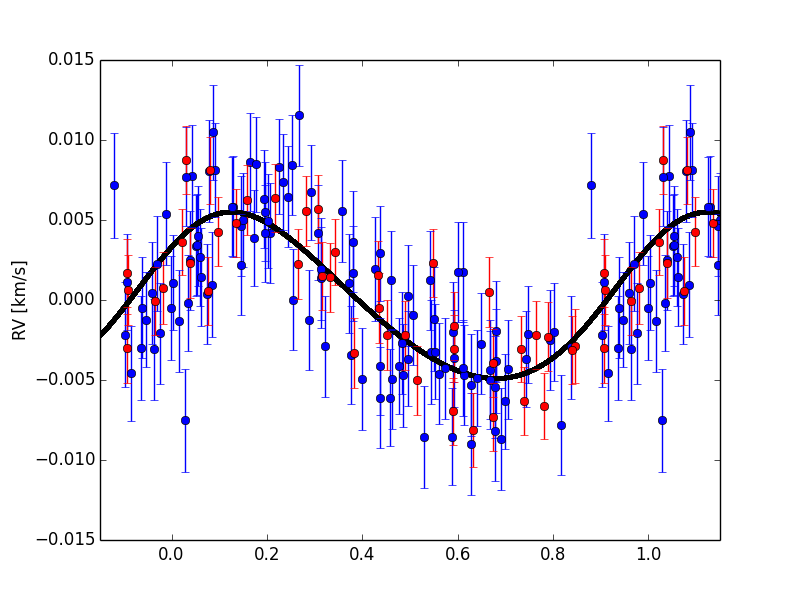}
   \includegraphics[width=\hsize]{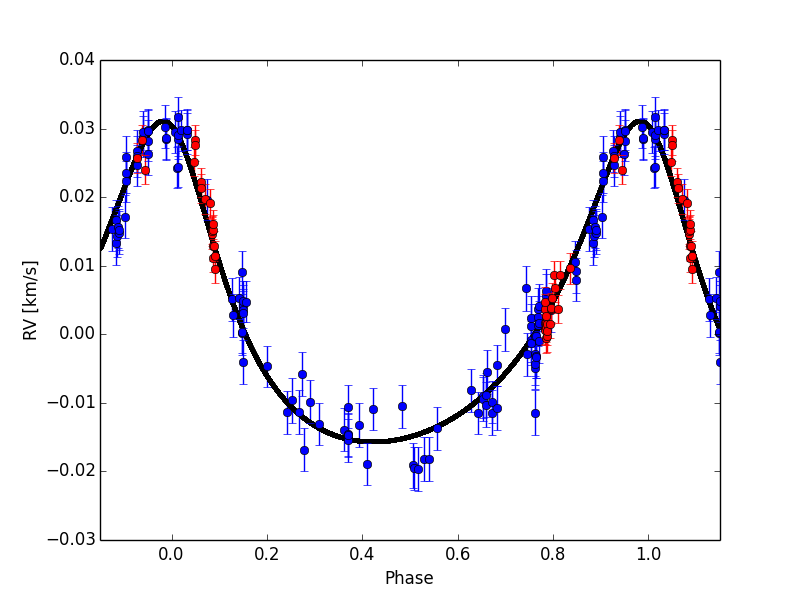}
      \caption{Phase-folded radial velocities of HD190360c (\textit{upper panel}) and HD190360b (\textit{bottom panel}) for HIRES (blue dots) and SOPHIE (red dots) with the best-fit orbital solution (solid line). Error bars include the photon noise and jitter.}
         \label{HD190360tot_phase}
\end{figure}

Finally, we combined our data set with the Keck observations used by \cite{Wright2009} to update orbital parameters for the HD190360 system. 100 MCMC chains of 300 000 steps were run with PASTIS to conduct a global search without a priori for a two-planet fit. The RV time series with our orbital solution is plotted in Fig. \ref{HD190360tot_orb}, and the phase-folded radial velocities for both HD190360 b and c are plotted in Fig. \ref{HD190360tot_phase}. The orbital and physical parameters of the two planets are reported in Table \ref{table:HD190360tot}.

The updated orbital parameters all agree well (within 1 $\sigma$) or marginally well (within 2 $\sigma$) with the values previously published in \cite{Wright2009}. The precision on all parameters is significantly improved except for the semi-amplitude and $\omega$. Assuming a stellar mass of $0.96 \pm 0.1$ {\Msun} as in \cite{Vogt2005}, the minimum mass of HD190360 b and c is $1.50 \pm 0.15$ {\Mjup} and $20.28 \pm 3.16$ M$_{\oplus}$, respectively.

The additional jitter (intrumental and stellar noise) necessary to set the reduced {\kid} to 1 is 2.9 {\ms} for HIRES and 1.8 {\ms} for SOPHIE. Additionally, the dispersion of the residuals is significantly higher for the HIRES data (3.3 {\ms}) than for SOPHIE (2.4 {\ms}). To take into account the much longer time span of the Keck observations (11.9 years), we computed the median dispersion of the residuals for all the 2.4 year windows in the HIRES data. Its value is 3.1 {\ms}, still significantly higher than for SOPHIE. This could be the result of either a higher instrumental noise or/and a higher stellar jitter at the time of the Keck observations. The latter case seems unlikely since the {\logrhk} reported in \cite{Vogt2005} is close to our value at -5.09.

\begin{table}
\caption{Updated orbital and physical parameters of HD190360 b and c}             
\label{table:HD190360tot}      
\centering                          
\begin{tabular}{c c c}        
\hline\hline                 
Parameter & HD190360b & HD190360c \\    
\hline                        
   P [days] & $2867.9 \pm 7.7$ & $17.1186 \pm 0.0016$\\      
   $T_0$ [BJD] & $59271 \pm 19$ & $55570.3^{+1.5}_{-2.9}$ \\
   $e$ & $0.343 \pm 0.017$ & $0.107 \pm 0.07$  \\
   $\omega$ [deg] & $14.7 \pm 32$  & $305.8^{+39}_{-280}$\\
   K [\ms] & $23.39 \pm 0.46$ & $5.20 \pm 0.37$\\
   msini [$M_\oplus$]& $475.16 \pm 49.0$  & $20.28 \pm 3.16$ \\  
\hline
   $N_{meas}$ (SOPHIE) & \multicolumn{2}{c}{41}\\
   $N_{meas}$ (HIRES) & \multicolumn{2}{c}{107}\\ 
   Jitter (SOPHIE) [\ms] & \multicolumn{2}{c}{1.8}\\\
   Jitter (HIRES) [\ms] & \multicolumn{2}{c}{2.9}\\
   $\sigma$ (O - C) (SOPHIE) [\ms] & \multicolumn{2}{c}{2.4}\\  
   $\sigma$ (O - C) (HIRES) [\ms] & \multicolumn{2}{c}{3.3}\\
\hline                                   
\end{tabular}
\end{table}

\section{HD164595 data analysis}

\subsection{Stellar properties}

HD164595 is a G2V star with a magnitude V=7.08. Its Hipparcos parallax ($\pi = 34.57 \pm 0.5$ mas) implies a distance of $28.93 \pm 0.4$ pc. This star is considered to be one of the closest known solar analogs. Its parameters were derived by \cite{Porto2014} in the context of their photometric and spectroscopic survey of solar twin stars within 50 parsecs of the Sun. From the {\sophie} CCFs and using the calibration given by \cite{Boisse2010}, we derived a {\vsini} of 2.1 $\pm$ 1 {\kms}. HD164595 is known to be a quiet star. \cite{Wright2004} reported a {\logrhk} of -5.0 from observations made between 1998 and 2003. More recently, \cite{Isaacson2010} reported a slight trend of {\logrhk} from -4.97 to -4.86 between 2005 and 2009. Following the method established by Boisse et al. (2010), we derived from our {\sophie} spectra an average value of -4.86 $\pm$ 0.05 for {\logrhk} . Furthermore, computing the median value for each season, we find that the {\logrhk} slightly increased from -4.91 in 2012 to -4.88 in 2013 and finally -4.81 in 2014.  We also note that the bisector span shows an increasing dispersion over three seasons from 1.7 {\ms} in 2012 to 2.0 {\ms} in 2013 and finally 3.3 {\ms} in 2014. During the last season, the bisector span is marginally correlated with the RV. All these factors seem to indicate that the stellar activity of HD164595 evolves along a magnetic cycle toward increased activity. Table \ref{table:stel} summarizes the stellar parameters of HD164595.

\begin{table}
\caption{Observed and inferred stellar parameters for HD164595.}             
\label{table:stel}      
\centering                          
\begin{tabular}{c c c}        
\hline\hline                 
Parameter & HD164595 & Reference \\    
\hline                        
   Spectral Type & G2V & Hipparcos\\      
   $\pi$ [mas] & $34.57 \pm 0.5$ & Hipparcos \\
   Distance [pc] & $28.93 \pm 0.4$ & Hipparcos \\
   V & 7.1 & Hipparcos \\
   B-V & 0.635 & \cite{Porto2014} \\
   $M_{\star}$ [\Msun] & $0.99 \pm 0.03$ & \cite{Porto2014} \\
   $T_{\textsl{eff}}$ [K] & $5790 \pm 40$ & \cite{Porto2014} \\
   $\log g$ & $4.44 \pm 0.05$ & \cite{Porto2014}\\
   $[Fe/H]$  & $-0.04 \pm 0.08$ & \cite{Porto2014} \\  
   $v \sin i$ [\kms] & $2.1 \pm 1$ & this paper \\
   $<\log (R'_{HK})>$ & $-4.86 \pm 0.05$ & this paper \\ 
\hline                                   
\end{tabular}
\end{table}

\subsection{Distinguishing systematic signals and aliases}

\begin{figure}
   \centering
   \includegraphics[width=\hsize]{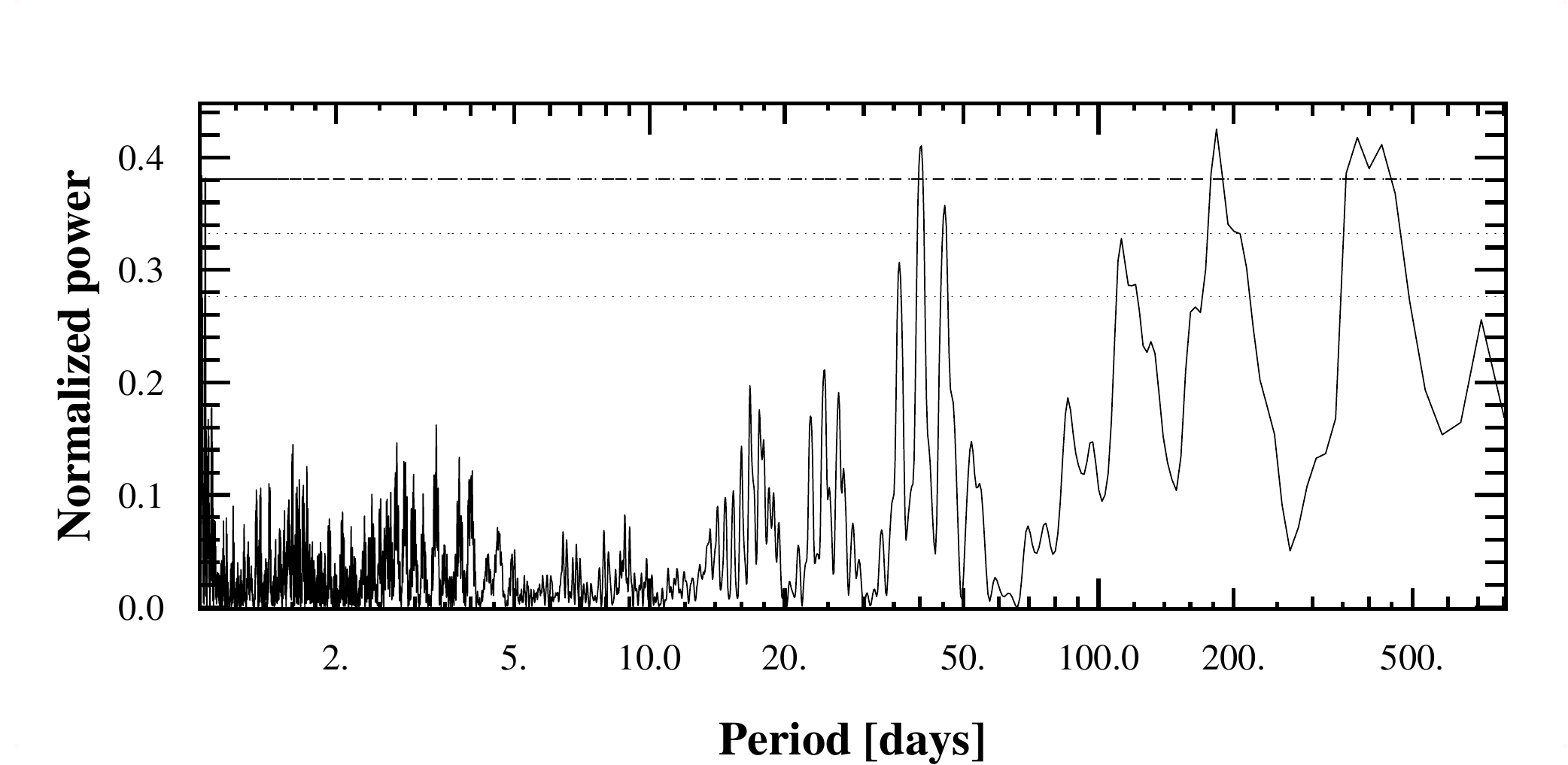}
   \includegraphics[width=\hsize]{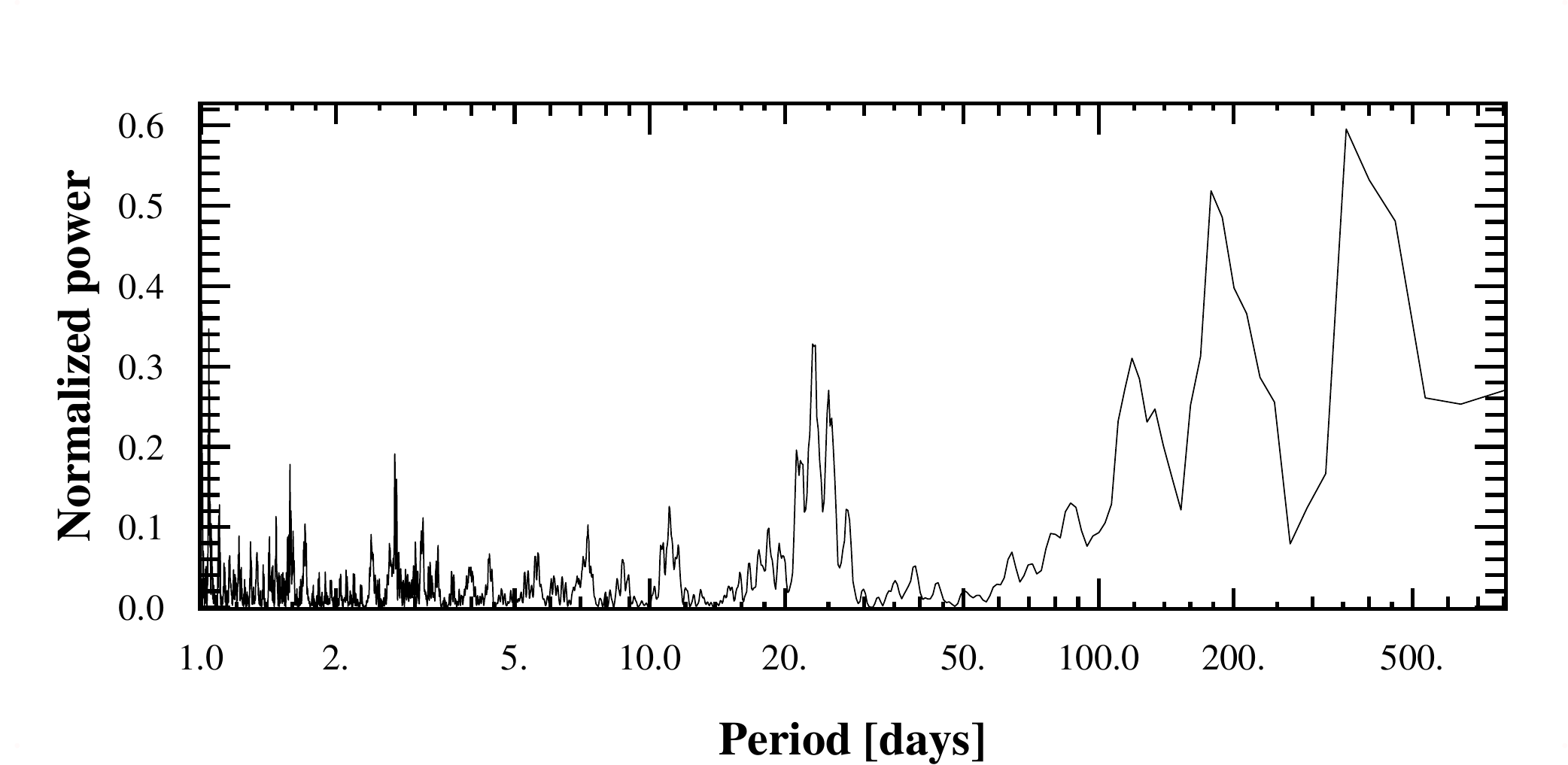}
   \includegraphics[width=\hsize]{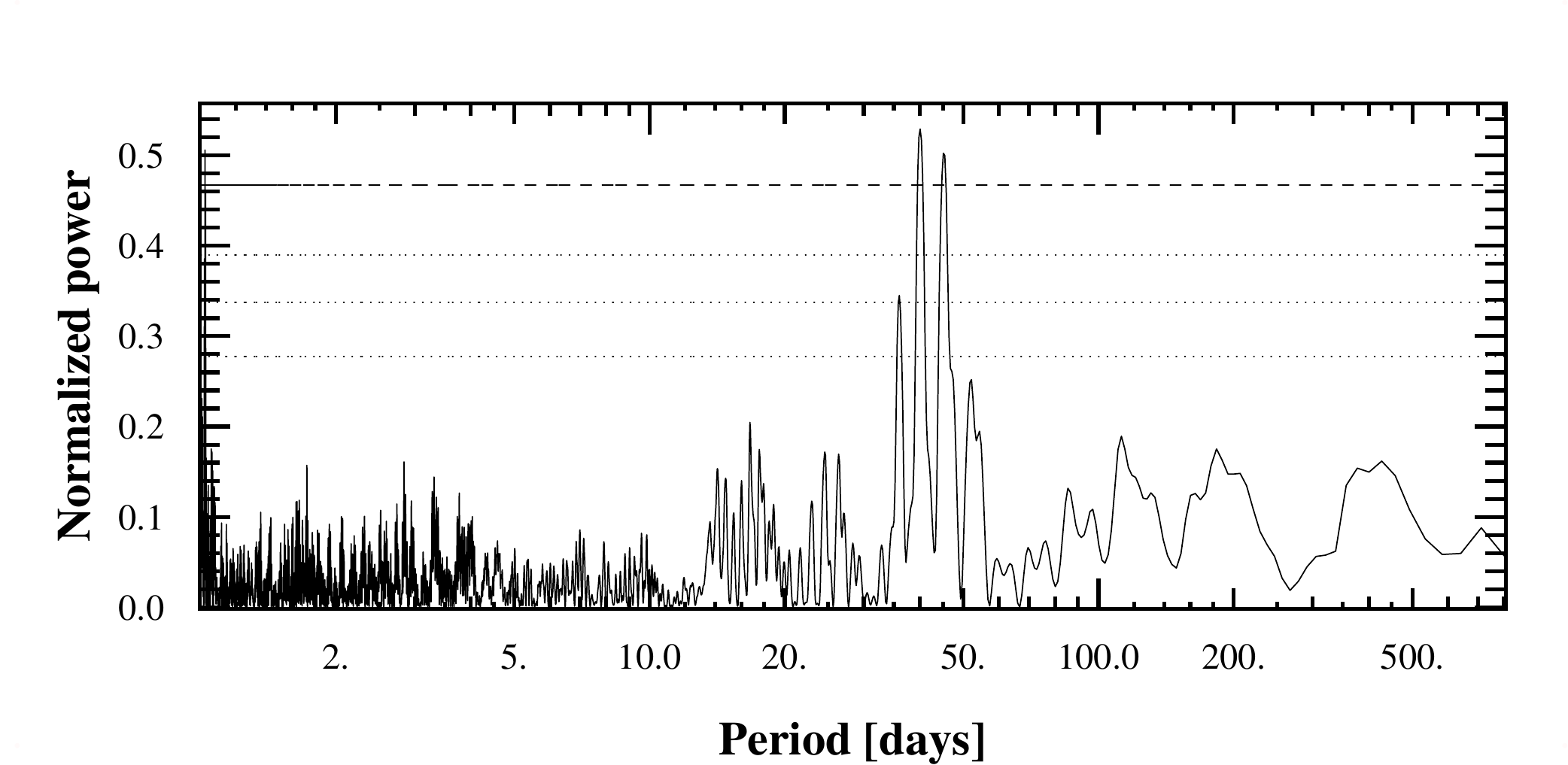}
      \caption{\textit{Upper panel} : Generalized Lomb-Scargle (GLS) periodogram of HD164595 RV before the constant master correction, with $10^{-3}$ FAP level (dashed line). \textit{Middle panel} : GLS periodogram of the zero-point drift correction applied to HD164595 dates of observation. \textit{Bottom panel} : GLS periodogram of the corrected HD164595 RV data, with $10^{-5}$ FAP level (dashed line).}
         \label{per_cor}
   \end{figure}

The Generalized Lomb-Scargle periodogram of the uncorrected radial velocities of HD164595 is displayed in the upper panel of Fig.~\ref{per_cor}. \textbf{A number of significant features appear, the most interesting being a triplet of peaks at 40$\pm$5 days.} The highest one lies below the $10^{-3}$ false-alarm probablity (FAP) level. The FAP was calculated with Yorbit \citep{Segransan2011} by scrambling the RV data many times and counting the number of randomly generated peaks in the corresponding periodograms with an amplitude higher than the peak of the original set. Here, the 40-day signal of the original data was systematically higher than the peaks of 10000 scrambled sets. Longer period signals are also found around 120, 180, and 400 days. They are
similar to the aliases of the one-year observational period of the spectral window. The middle panel of Fig.~\ref{per_cor} shows the periodogram of the RV constant master sampled at the same observing dates as HD164595. We note that except for the 40-day triplet, other features are the same in the two periodograms. After correction of the systematic drift, that is, after subtracting the constant master RVs from the RVs of HD164595, the only significant feature in the periodogram (Fig \ref{per_cor}, bottom panel) is the 40-day triplet with a FAP lower than $10^{-5}$ . This signal is then clearly not introduced by our correction, while other peaks were successfully reduced or removed.

The multiplicity of the 40-day feature is due to the convolution with the one-year observational period of the spectral window. The automatic de-aliasing algorithm CLEAN \citep{Roberts1987} implemented in Yorbit is another way to confirm the unicity of the physical signal in the feature by cleaning up the periodogram. While giving an important diagnostic, this method cannot be used to determinate which frequency is the physical one. This problem is particularly essential for the 40-day and 45-day solutions, which are equally satisfying (with similar amplitudes in the GLS periodogram and a  similar $\chi2$ and residual dispersion for their respective fits). 

\begin{figure}
   \centering
   \includegraphics[width=\hsize]{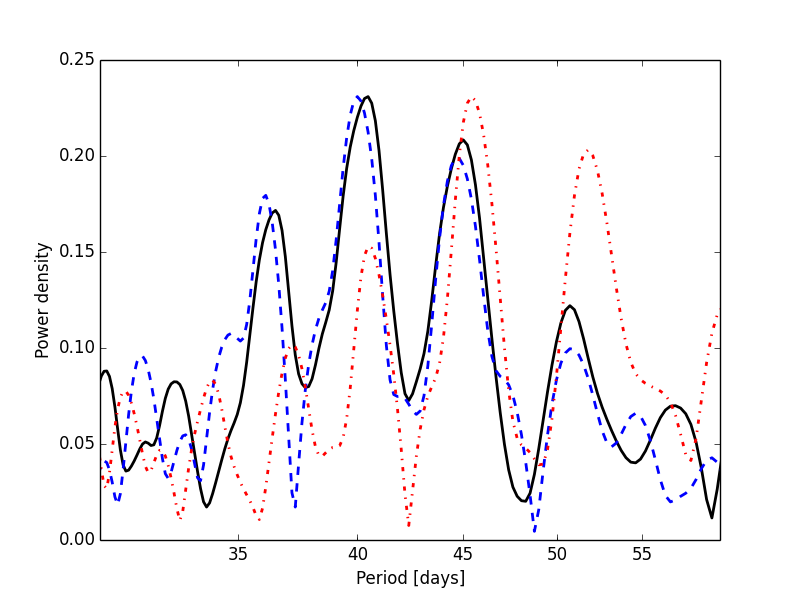}
      \caption{GLS periodogram of the complete HD164595 RV data (with all the 75 measurements) (solid black curve), the 40-day sinusoid (dashed blue curve) and the 45-day sinusoid (dotted and dashed red curve). The 40-day sinusoid clearly is the best match to the RV data.}
         \label{per_alias}
   \end{figure}

\cite{Dawson2010} proposed a method for solving this problem. The concept is to simulate a sinusoidal signal at the period and amplitude of each observational alias and to sample it at the date of observations. By comparing their periodograms, the one that significantly better matches the data periodogram corresponds to the true physical frequency. If the test is inconclusive, it means that the noise prevents identification, and new observations are required. This was the case when we apply this method using only our 66 high-quality observations. We then used our complete data set of 75 measurements (see Fig.~\ref{per_alias}) and determined that the physical frequency is the 40-day peak. 

\subsection{Distinguishing stellar activity}

\begin{figure}
   \centering
   \includegraphics[width=\hsize]{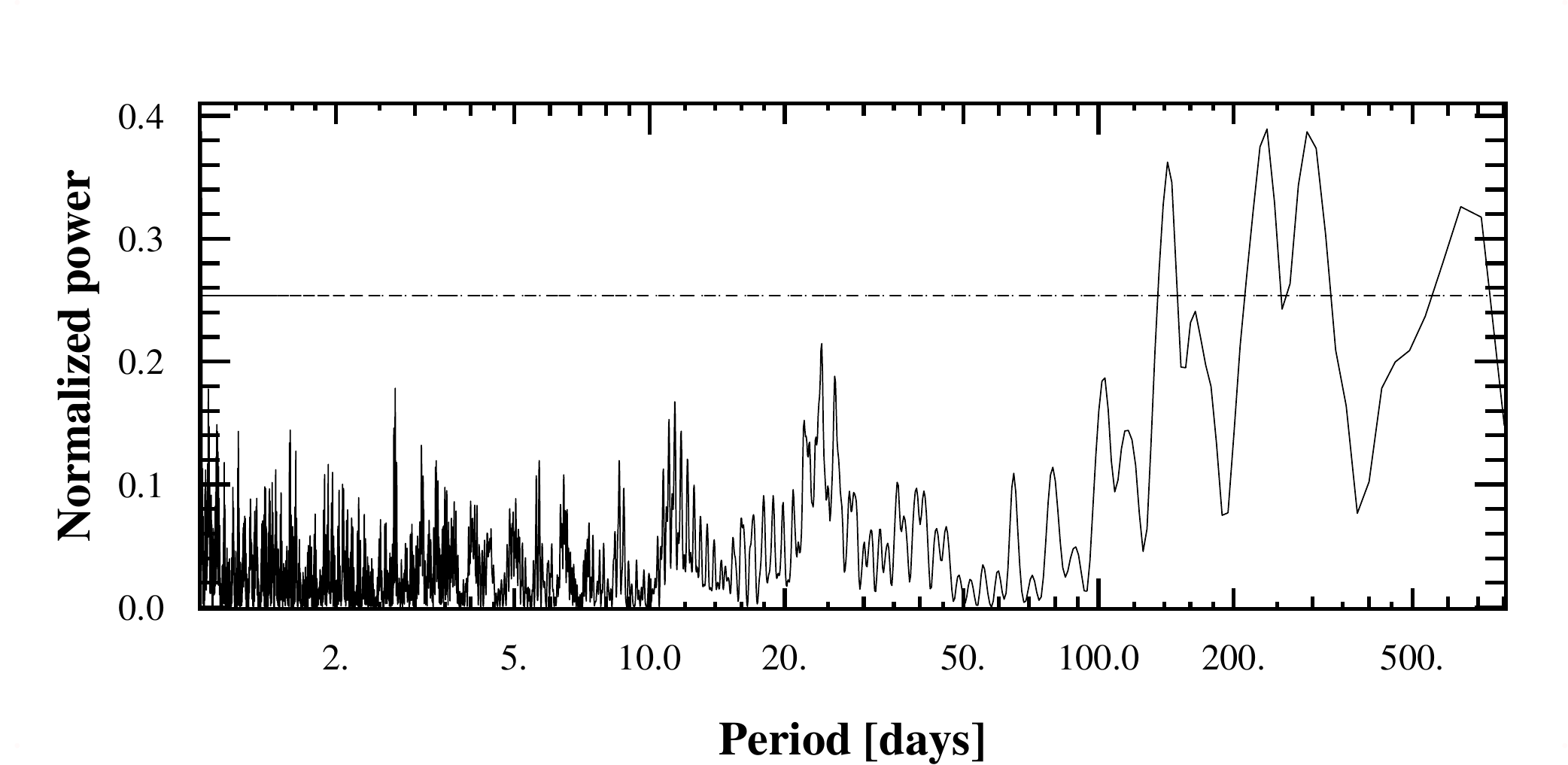}
      \caption{HD164595 bisector Generalized Lomb-Scargle periodogram with 0.1 FAP level (dashed line). Long-period signals above this level are caused by a trend in the last season that is marginally correlated with the radial velocities.}
         \label{bisector}
\end{figure}

We studied the possibility that the 40-day signal might be due to star-induced activity and correspond to the rotational period. While we noted a slight increase in activity level with time, HD164595 remained a relatively quiet star even in the last observing season with a {\logrhk} below -4.80. Figure \ref{bisector} shows the periodogram of the bisector span, which can be used to reflect stellar activity. No signal is present at 40 days or at its first harmonics. This also remains true with the FWHM and the {\logrhk}. There is no correlation between RV and these activity indicators
either, except for the already mentioned marginal correlation between RV and the bisector, but only in the last season. Moreover, the rotational period of HD164595 estimated from the {\logrhk} using the calibration published by \cite{Mamajek2008} is $20 \pm 5$ days, which is quite distinct from the  observed 40-day signal. We note that no 20-day signal appears in the RV periodogram,
which means that the 40-day signal cannot be an observational alias. The HIPPARCOS photometry does not show any significant variations at a level of 13 mmag (with 147 points), and with no hint of periodic signal in the periodogram. We also checked that the 40-day signal maintains the same phase and amplitude over the last two seasons when fitted independantly (the first season does not have enough data points to conclude). We are therefore confident that the 40-day signal is not induced by activity and is planetary in origin.

\subsection{Radial-velocity analysis and orbital solution}

\begin{figure}
   \centering
   \includegraphics[width=\hsize]{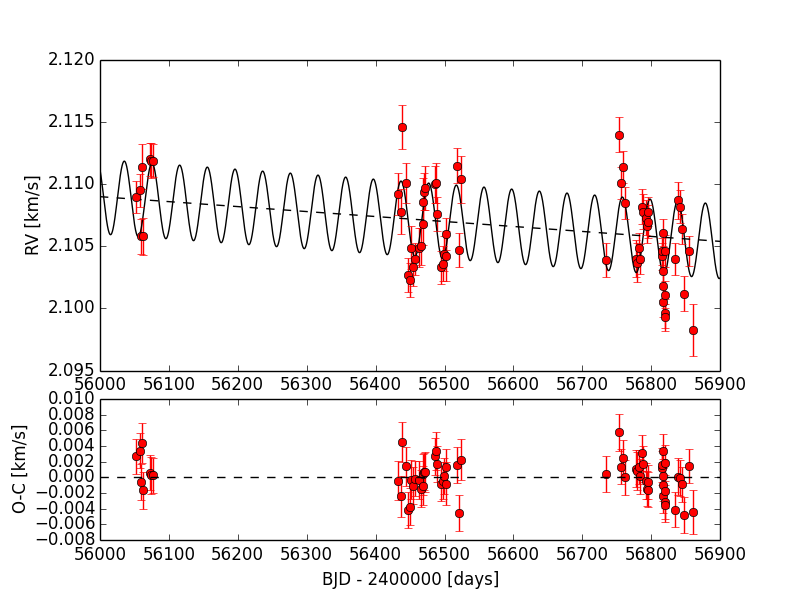}
   \includegraphics[width=\hsize]{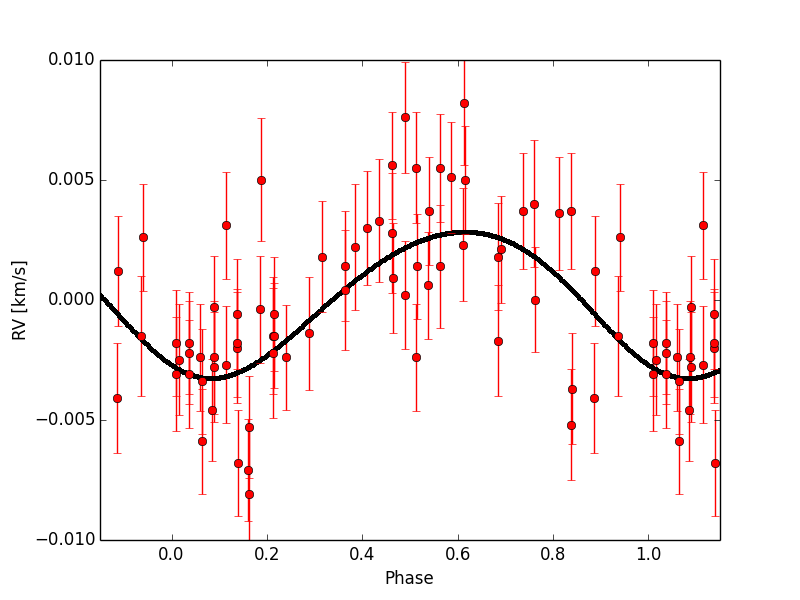}
      \caption{\textit{Upper panel} : Radial velocities of HD164595 with the best-fit orbital solution (solid line). Error bars include the photon and instrumental noise and the stellar jitter. \textit{Bottom panel} : Phase-folded radial velocities of HD164595 with the best-fit orbital solution (solid line). Error bars include the photon and instrumental noise and the stellar jitter.}
         \label{HD164595_orb}
\end{figure}

For the rest of the analysis and the fit of a Keplerian signal we only kept the high-quality data set (66 RV measurements) after removing the nine low-quality measurements (low S/N, moon light contamination, anormal simultaneous thorium-argon flux, see Sect. 2). The RMS of the RV time series shown in Fig.~\ref{HD164595_orb} (upper panel) is 3.73 {\ms}. A slight long-term trend is visible. We used PASTIS to fit a Keplerian and a linear drift to our data without a priori. We ran 20 MCMC chains with 300 000 steps each, and discarded the solutions that did not converge to the 40-day period (some converged to the observational aliases, in particular the 45-day period).  
        
Table \ref{table:planet} reports the orbital parameters and their uncertainties for a\ Keplerian model with a linear trend, and the best fit is plotted in Fig. \ref{HD164595_orb}. The fitted model of the planet has a $40-$ day period and a semi-amplitude of $3.05 \pm 0.41$ {\ms} with an insignificant eccentricity. Assuming a mass of $0.99 \pm 0.03$ {\Msun} for the star \citep{Porto2014}, the minimum mass of the planet is $16.1 \pm 2.7$ M$_{\oplus}$. With a jitter of 1.8 {\ms}, the reduced {\kid} is equal to 1. The residuals after the fit present a dispersion of 2.3 {\ms} , and no significant feature appears in the corresponding periodogram. An interesting trend can be observed in the last season. It might correspond to either the activity-related drift that is correlated with the bisector span, or to a shift in the global trend and a subsequent incomplete correction. Trying to correct this trend on the data does not significantly change our reported solution. 

Another analysis using the genetic algorithm built into Yorbit \citep{Segransan2011} yielded nearly identical parameter values and errors (P = $39.99\pm0.19$ days, K=$2.96\pm0.40$ {\ms} and e=$0.078\pm0.14$). We note that the analysis of the uncorrected data does not change our results, but only increases the error bars of our orbital solution, the scatter of the residuals (from 2.2 to 2.9 {\ms}), and the corresponding jitter (from 1.8 to 2.7 {\ms}).

\begin{table}
\caption{Orbital and physical parameters of HD164595b}             
\label{table:planet}      
\centering                          
\begin{tabular}{c c}        
\hline\hline                 
Parameter & HD164595b \\    
\hline                        
   P [days] & $40.00 \pm 0.24$ \\      
   $T_0$ [BJD] & $56280 \pm 12$  \\
   $e$ & $0.088^{+0.12}_{-0.066}$  \\
   $\omega$ [deg] & $145^{+160}_{-110}$  \\
   K [\ms] & $3.05 \pm 0.41$ \\
   msini [$M_\oplus$] & $16.14 \pm 2.72$ \\
   $a$ [AU]  & 0.23 \\  
   \textbf{drift [m/s/yr]} & $-2.34 \pm 0.44$  \\
\hline
   $N_{meas}$ & 66 \\      
   Data span [days] & 809  \\
   $\sigma$ (O - C) [\ms] & 2.3 \\
   Jitter [\ms] & 1.8   \\    
\hline                                   
\end{tabular}
\end{table}

\section{Discussion and conclusion}

We have reported the discovery of HD164595b with the {\sophie} spectrograph. This source is a warm Neptune-like exoplanet with a minimum mass of $16.1 \pm 2.7$ M$_{\oplus}$ orbiting a bright solar-analog star with a period of $40.00 \pm 0.24$ days with no significant eccentricity. An additional long-term RV drift is seen on our data. At this stage, we cannot conclude on its origin. It might be due to a long-period additional companion in the system, but it might likewise be the signature of a stellar magnetic cycle since we observe an slight increase of the activity index over our observing span of 2.2 years. According to \cite{Lovis2011b}, a change of 0.1 dex in the {\logrhk} may introduce a change of
up to 10 {\ms}  in radial velocity. At this stage, we cannot completely exclude an under-correction of the zero-point drift
either. Additional {\sophie} measurements will help to conclude on the long-term drift. 

To validate and illustrate the capacity of {\sophie} to reach the 2 {\ms} precision thanks to an appropriate 
correction of systematic instrumental effect, we refined the orbital parameters of the known 
planetary system HD190360 based on {\sophie} and HIRES radial velocity measurements. With this precision on all timescales, {\sophie} can detect exoplanets with an RV semi-amplitude as low as 3 {\ms}. 
Nevertheless, pushing the RV precision down to 1 {\ms} will be crucial to efficiently cover the parameter space of low-mass exoplanets. Additional improvements are ongoing on {\sophie,
such as}  the thermal control of the spectrograph tripod and the implementation of a Fabry-Perot etalon within the calibration unit to improve the determination of the simultaneous drift. 

\begin{figure}
   \centering
   \includegraphics[width=\hsize]{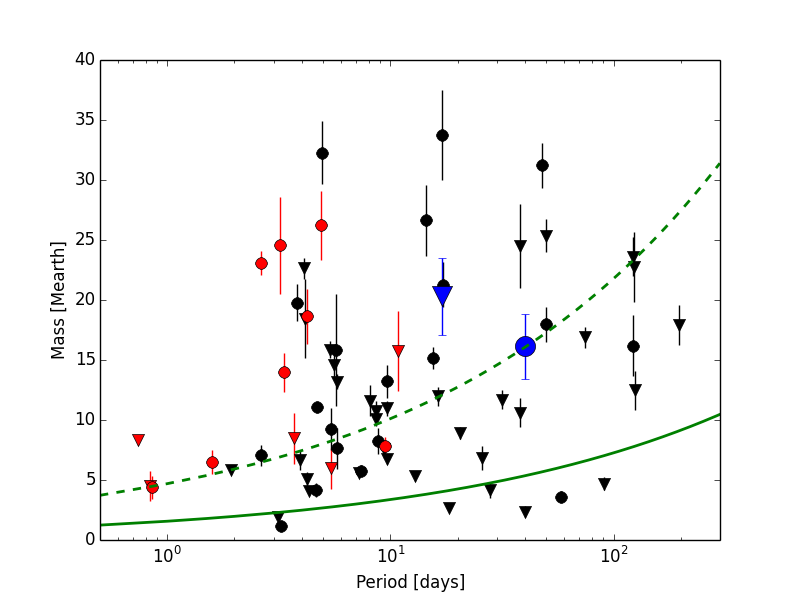}
      \caption{Mass - orbital period of known planets with $m\sini<40$M$_{\oplus}$, $\Delta m\sin i / m\sin i < 0.3$. Another three planets with $P>300$ days are not represented because of visibility problems. Transiting planets are shown in red, while the others are black. Triangles correspond to planets in multiplanetary systems. HD164595b and HD190360c are the blue circle and triangle. The green curves refer to the 3 \ms (dashed) and 1 \ms (solid) iso RV semi-amplitudes for a solar-type star. This diagram is based on data from the exoplanet.org database.}
         \label{massor}
\end{figure}

About 70 exoplanets with m$\sini$ smaller than 40 M$_{\oplus}$ and an  uncertainty on m$\sini$ smaller than 30\% are known. They are displayed in the mass - orbital period diagram in Fig. \ref{massor}. In that sample, no transiting low-mass planets with period greater than ten days are known except for Kepler-20c \citep{Gautier2012}, which has a period  of 10.9 days. Two-thirds
of the 18 exoplanets with orbital periods longer than ten days
are part of a multiple planetary system. Only three transiting planets with masses in the range 10-20 M$_{\oplus}$ are known (Kepler-20c, GJ3470b, and HAT-P-26b). Their radii range from 3.1 to 6.3 R$_{\oplus}$ , illustrating the diversity in density of this type of objects. This also underlines the current lack of constraints on these objects. The transition between super-Earths and Neptune-like planets and its dependance on the orbital period is poorly understood. The actual number of planets with precise density measurements is indeed far too low to cover the parameter space and constrain interior, formation, or evolution models for the low-mass planet population. The future transit search missions around bright stars TESS \citep{Ricker2014} and CHEOPS \citep{Broeg2013,Fortier2014} will address this issue from two complementary perspectives. While TESS will substantially increase the number of transiting exoplanets suitable for high-precision RV measurement, CHEOPS will mesure precise radii of known exoplanets that have been discovered with radial velocities to derive precise densities. However, the low overall transit probability of the non-transiting planets of this sample ($\sim$300\%) implies that CHEOPS will need a significant number of targets at various periods, requiring continuous efforts on low-mass RV surveys.   

In this context, HD164595b is an interesting and potential target for the follow-up transit mission CHEOPS, although its transit probability is only 2\%. HD164595 matches all CHEOPS requirements in terms of coordinates, magnitude, and spectral type. The uncertainty of the transit window extrapolated to 2018 is about six days at this stage, but long-term monitoring of this target with {\sophie} is planned to reduce this uncertainty. Following the mass-radius relation given by \cite{Marcy2014} and assuming that $\sini$=1, the expected size of HD164595b is around 4.7 R$_{\oplus}$. Hence the expected transit probably has a depth close to 1.9 mmag with a duration of 6.1 hours. We note, however, that the three transiting planets with similar masses as HD164595b that were previously mentioned have a radius between 3.1 and 6.3 R$_{\oplus}$. In case of transit, CHEOPS will be fully appropriate to derive the radius with a precision better than 10\%. Finally, CHEOPS could also be used to detect transits of additional short-period low-mass companions that are unseen in radial velocities in the HD164595 system.

\begin{acknowledgements}
     We gratefully acknowledge the Programme National de Plan{\'e}tologie (telescope time attribution and financial support) of CNRS/INSU, the Swiss National Science Foundation, and the Agence Nationale de la Recherche (grant ANR-08-JCJC-0102-01) for their support. We warmly thank the OHP staff for their great care in optimizing the observations. N.C.S. acknowledges support by Fundação para a Ciência e a Tecnologia (FCT) through Investigador FCT contract of reference IF/00169/2012 and POPH/FSE (EC) by FEDER funding through the program “Programa Operacional de Factores de Competitividade - COMPETE”. We further acknowledge FCT through FEDER funds in program COMPETE, as well as through national funds, in the form of the grants with reference RECI/FISAST /0176/2012 (FCOMP-01-0124-FEDER-027493), RECI/FIS-AST/0163/2012 (FCOMP-01-0124-FEDER-027492), and UID/FIS/04434/2013. A.S. is supported by the European Union under a Marie Curie Intra-European Fellowship for Career Development with reference FP7-PEOPLE-2013-IEF, number 627202. P.A.W acknowledges the support of the French Agence Nationale de la Recherche (ANR), under program ANR-12-BS05-0012 "Exo-Atmos”.
\end{acknowledgements}
\begin{longtab}
\begin{longtable}{l c c c}        
\caption{\label{table:obs} Radial-velocity measurements, error bars and bissector for HD164595. 
The dates with an asterix correspond to low quality data discarded for the analysis.}\\               
\hline                 
BJD & RV [$\kms$] & Uncertainty [$\kms$] & Biss [$\kms$] \\    
\hline                        
56051.596 & 2.1088 & 0.0013 & -0.0248 \\
56058.576 & 2.1094 & 0.0013 & -0.0240 \\
56059.487 & 2.1056 & 0.0014 & -0.0282 \\
56061.524 & 2.1112 & 0.0018 & -0.0260 \\
56062.557 & 2.1057 & 0.0015 & -0.0285 \\
56072.524 & 2.1118 & 0.0013 & -0.0262 \\
56074.524 & 2.1118 & 0.0014 & -0.0283 \\
56076.503 & 2.1117 & 0.0013 & -0.0295 \\
56432.514 & 2.1091 & 0.0017 & -0.0270 \\
56434.581* & 2.1123 & 0.0014 & -0.0270 \\
56436.544 & 2.1077 & 0.0018 & -0.0240 \\
56438.527 & 2.1145 & 0.0018 & -0.0238 \\
56443.518 & 2.1100 & 0.0016 & -0.0288 \\
56447.595 & 2.1026 & 0.0014 & -0.0253 \\
56449.414 & 2.1022 & 0.0014 & -0.0265 \\
56451.437 & 2.1048 & 0.0017 & -0.0273 \\
56454.408 & 2.1032 & 0.0015 & -0.0262 \\
56456.381 & 2.1039 & 0.0013 & -0.0290 \\
56462.520 & 2.1048 & 0.0016 & -0.0262 \\
56465.499 & 2.1049 & 0.0015 & -0.0277 \\
56468.582 & 2.1067 & 0.0017 & -0.0280 \\
56469.377 & 2.1085 & 0.0019 & -0.0290 \\
56470.421 & 2.1093 & 0.0015 & -0.0247 \\
56471.402 & 2.1095 & 0.0018 & -0.0238 \\
56486.475 & 2.1099 & 0.0015 & -0.0290 \\
56487.540 & 2.1099 & 0.0016 & -0.0278 \\
56489.510 & 2.1075 & 0.0014 & -0.0258 \\
56495.437 & 2.1032 & 0.0013 & -0.0280 \\
56497.549 & 2.1034 & 0.0014 & -0.0303 \\
56499.486 & 2.1042 & 0.0014 & -0.0267 \\
56501.441 & 2.1059 & 0.0013 & -0.0262 \\
56502.480 & 2.1041 & 0.0020 & -0.0327 \\
56517.400 & 2.1114 & 0.0014 & -0.0252 \\
56520.402* & 2.1097 & 0.0013 & -0.0283 \\
56521.390 & 2.1045 & 0.0014 & -0.0270 \\
56524.431 & 2.1102 & 0.0019 & -0.0288 \\
56730.676* & 2.1143 & 0.0013 & -0.0347 \\
56733.663* & 2.1124 & 0.0017 & -0.0248 \\
56734.647 & 2.1037 & 0.0014 & -0.0280 \\
56753.586 & 2.1137 & 0.0014 & -0.0258 \\
56755.611 & 2.1098 & 0.0013 & -0.0222 \\
56756.578* & 2.1169 & 0.0023 & -0.0287 \\
56758.605 & 2.1112 & 0.0013 & -0.0273 \\
56761.624 & 2.1082 & 0.0013 & -0.0240 \\
56777.522 & 2.1037 & 0.0015 & -0.0283 \\
56778.570 & 2.1034 & 0.0016 & -0.0200 \\
56782.561 & 2.1047 & 0.0012 & -0.0298 \\
56783.568 & 2.1039 & 0.0012 & -0.0313 \\
56786.585 & 2.1081 & 0.0014 & -0.0287 \\
56788.581 & 2.1076 & 0.0014 & -0.0267 \\
56792.607 & 2.1072 & 0.0014 & -0.0288 \\
56793.555 & 2.1065 & 0.0013 & -0.0295 \\
56794.603 & 2.1078 & 0.0012 & -0.0313 \\
56795.529 & 2.1070 & 0.0013 & -0.0342 \\
56815.447 & 2.1043 & 0.0011 & -0.0312 \\
56815.483 & 2.1046 & 0.0011 & -0.0323 \\
56816.560 & 2.1007 & 0.0012 & -0.0323 \\
56816.592 & 2.1031 & 0.0012 & -0.0327 \\
56817.421 & 2.1020 & 0.0011 & -0.0323 \\
56817.593 & 2.1062 & 0.0011 & -0.0282 \\
56819.464 & 2.1047 & 0.0013 & -0.0322 \\
56819.564 & 2.0997 & 0.0012 & -0.0332 \\
56820.440 & 2.0995 & 0.0011 & -0.0310 \\
56820.514 & 2.1013 & 0.0011 & -0.0335 \\
56834.466 & 2.1039 & 0.0013 & -0.0295 \\
56838.411 & 2.1085 & 0.0015 & -0.0278 \\
56841.415 & 2.1080 & 0.0013 & -0.0322 \\
56844.519 & 2.1062 & 0.0012 & -0.0313 \\
56847.487 & 2.1010 & 0.0014 & -0.0320 \\
56850.546* & 2.0970 & 0.0016 & -0.0363 \\
56854.401 & 2.1044 & 0.0012 & -0.0315 \\
56860.493 & 2.0982 & 0.0021 & -0.0422 \\
56900.431* & 2.0949 & 0.0012 & -0.0310 \\
56901.312* & 2.1002 & 0.0013 & -0.0317 \\
56902.312* & 2.0951 & 0.0015 & -0.0360 \\
\hline                                   
\end{longtable}
\end{longtab}

\begin{longtab}
\begin{longtable}{c c c}        
\caption{\label{table:obsHD190360} Radial-velocity measurements and error bars for HD190360 (after correction).}\\      
\hline                 
BJD & RV [$\kms$] & Uncertainty [$\kms$]\\    
\hline  
55756.500 & -45.2113 & 0.0013 \\
55757.487 & -45.2097 & 0.0012 \\
55758.507 & -45.2077 & 0.0011 \\
55759.532 & -45.2057 & 0.0012 \\
55760.522 & -45.2055 & 0.0011 \\
55761.349 & -45.2096 & 0.0012 \\
55762.493 & -45.2104 & 0.0011 \\
55763.380 & -45.2151 & 0.0012 \\
55765.622 & -45.2167 & 0.0011 \\
55767.627 & -45.2197 & 0.0014 \\
55769.479 & -45.2178 & 0.0011 \\
55772.359 & -45.2144 & 0.0011 \\
55785.487 & -45.2181 & 0.0011 \\
55793.383 & -45.2056 & 0.0011 \\
55796.467 & -45.2088 & 0.0011 \\
55798.493 & -45.2107 & 0.0012 \\
55813.428 & -45.2038 & 0.0011 \\
55818.330 & -45.2123 & 0.0010 \\
55835.417 & -45.2154 & 0.0010 \\
55848.299 & -45.2048 & 0.0010 \\
55908.236 & -45.2075 & 0.0014 \\
56175.409 & -45.1906 & 0.0012 \\
56200.333 & -45.1865 & 0.0012 \\
56220.319 & -45.1856 & 0.0011 \\
56517.451 & -45.1907 & 0.0011 \\
56519.428 & -45.1901 & 0.0011 \\
56521.443 & -45.1921 & 0.0012 \\
56554.416 & -45.1969 & 0.0011 \\
56557.378 & -45.1977 & 0.0011 \\
56583.355 & -45.1921 & 0.0012 \\
56613.268 & -45.1914 & 0.0011 \\
56624.292 & -45.2050 & 0.0011 \\
56625.297 & -45.2042 & 0.0010 \\
56626.280 & -45.2035 & 0.0011 \\
56628.290 & -45.1997 & 0.0011 \\
56629.263 & -45.2015 & 0.0011 \\
56630.238 & -45.1979 & 0.0010 \\
56631.246 & -45.1935 & 0.0010 \\
56638.266 & -45.2044 & 0.0011 \\
56643.254 & -45.2092 & 0.0010 \\
56644.242 & -45.2058 & 0.0011 \\
\hline                                   
\end{longtable}
\end{longtab}




\end{document}